# CO oxidation by $Pt_2/Fe_3O_4$: metastable dimer and support configurations facilitate lattice oxygen extraction

**Authors:** Matthias Meier[1,2†], Jan Hulva[1†], Zdenek Jakub[1], Florian Kraushofer[1], Mislav Bobić[1], Roland Bliem[1‡], Martin Setvin[1,3], Michael Schmid[1], Ulrike Diebold[1], Cesare Franchini[2,4] and Gareth S. Parkinson[1]

[1]Institute of Applied Physics, TU Wien, Vienna, Austria.

[2]Computational Materials Physics, University of Vienna, Vienna, Austria.

[3]Department of Surface and Plasma Science, Faculty of Mathematics and Physics, Charles University, Prague, Czech Republic

[4]Alma Mater Studiorum – Università di Bologna, Bologna, Italy.

[‡]Current address: Advanced Research Center for Nanolithography, Science Park 106, 1098XG Amsterdam, The Netherlands

*Correspondence to: parkinson@iap.tuwien.ac.at

†These authors contributed equally to this work

**Abstract: Heterogeneous catalysts based on sub-nanometer metal clusters often exhibit strongly size-dependent properties, and the addition or removal of a single atom can make all the difference. Identifying the most active species and deciphering the reaction mechanism is extremely difficult, however, because it is often not clear how the catalyst evolves *in operando*. Here, we utilize a combination of atomically resolved scanning probe microscopies, spectroscopic techniques, and density functional theory (DFT)-based calculations to study CO oxidation by a model $Pt/Fe_3O_4(001)$ "single-atom" catalyst. We demonstrate that $(PtCO)_2$ dimers, formed dynamically through the agglomeration of mobile Pt-carbonyl species, catalyse a reaction involving the oxide support to form $CO_2$. $Pt_2$ dimers produce one $CO_2$ molecule before falling apart into two adatoms, releasing the second CO. Interestingly, $O_{lattice}$ extraction only becomes facile when both the Pt-dimer and the $Fe_3O_4$ support can access metastable configurations, suggesting that substantial, concerted rearrangements of both cluster and support must be considered for reactions occurring at elevated temperature.**

**125 character teaser:** Non-equilibrium cluster and support morphologies are important when modelling heterogeneous catalysis at elevated temperatures

The continuing trend to downsize the precious-metal component of supported heterogeneous catalysts has seen attention turn to the sub-nano regime (*1-9*). Here, supported clusters no longer resemble larger nanoparticles in either physical or electronic structure, and simple scaling laws no longer apply (*10*). Experiments utilizing size-selected clusters have clearly shown that the optimum particle size varies from reaction to reaction and system to system, and in some cases the addition or removal of just one atom can have a dramatic effect (*3, 9, 11*). Isolated atoms have been proposed to be catalytically active for some reactions (*12-18*), and so-called single-atom catalysis (SAC) has gained much attention as a bridge to well-understood, highly-selective homogeneous catalysts (*19-21*). Nevertheless, the field remains controversial because characterizing single-atom catalysts pushes the limits of current experimental techniques, and there remains much discussion as to whether catalytic activity really stems from isolated adatoms, or sub-nano particles (*1, 22-25*).

One of the biggest challenges to understanding such systems is that catalysts typically evolve under reaction conditions (*26*). Thus, a catalyst that begins life as a "single-atom" system, for example, can undergo processes that lead to a distribution of cluster sizes over time (*7, 27*), and any of the resulting clusters might be responsible for a high activity. Then there is the question of mechanism. Most fundamental SAC studies to date have utilized CO oxidation as a probe reaction, and Mars-van Krevelen (MvK) (*12, 13, 28*) and Eley-Rideal (*29, 30*) mechanisms have been proposed. Given the uncertainty around the structure of "real" single-atom catalysts, studies based on precisely-defined model systems (*27, 31-35*) are important to conclusively determine whether single atoms are catalytically active, and if so, how they work.

In this paper, we use a combination of atomically-resolved scanning-probe microscopy, surface-sensitive spectroscopy, and density functional theory to study CO oxidation on a Pt/$Fe_3O_4$(001) model catalyst. We show that (PtCO)$_2$ dimers are formed dynamically due to CO-induced sintering, and that these species catalyse CO oxidation through a reaction with the support. Characterizing the initial state by noncontact atomic force microscopy (ncAFM) confirms the (PtCO)$_2$ geometry determined by density functional theory (DFT) calculations, and allows direct imaging of individual CO molecules adsorbed on a sub-nano cluster. We demonstrate that CO oxidation occurs from a metastable (PtCO)$_2$ configuration that becomes available at elevated temperature, and that a rearrangement of the lattice of the support is also required to quantitatively reproduce the experimental results.

## Results

**Scanning Tunnelling Microscopy (STM) Measurements of Pt/Fe$_3$O$_4$(001).** The experiments described here rely on the remarkable stability of metal adatoms on the Fe$_3$O$_4$(001)-($\sqrt{2}\times\sqrt{2}$)R45° support[36, 37]. After UHV preparation, constant-current STM images of this surface exhibit rows of Fe atoms running in the <110> directions due to a termination at the Fe$_{oct}$-O plane of the inverse-spinel structure [36]. Surface O atoms are not imaged in STM because they have no density of states near the Fermi level (E$_F$). The ($\sqrt{2}\times\sqrt{2}$)R45° periodicity is linked to an ordered array of subsurface cation vacancies and interstitials in the subsurface layers [36]. Pt$_1$ adatoms bind strongly to two particular surface oxygen atoms within a surface unit cell (those without a subsurface Fe$_{tet}$ neighbour, labelled O1 in top view in Fig 1D), and the isolated Pt$_1$ atoms remain stable in this configuration to temperatures as high as 700 K in UHV. In previous work [38, 39], we have shown that the room-temperature adsorption of CO results in mobile Pt carbonyl species that rapidly agglomerate. Fig. 1A shows an STM image acquired at 78 K of a Pt/Fe$_3$O$_4$(001) model catalyst prepared by exposing 0.2 ML Pt$_1$ adatoms to 8×10$^{-5}$ mbar.s CO at 300 K. Here, 1 ML is defined as the density of possible Pt$_1$ adsorption sites, which is 1.42×10$^{14}$ cm$^{-2}$. The size of the resulting Pt clusters is assigned on the basis of experiments in which the CO-induced agglomeration was followed atom-by-atom with the STM, as shown previously [39]. The most common species appears as a double protrusion (indicated by a white arrow) between the rows of the underlying Fe$_3$O$_4$(001) support, and contains two Pt atoms. The axis of the protrusions is slightly rotated with respect to the rows. A cluster (red arrow) with an apparent height of 3.9 Å contains 5-6 Pt atoms. Adsorbed CO is not visible in STM images, but its presence on the clusters can be inferred from a peak at 288.7 eV in C1s X-ray photoelectron spectroscopy (XPS) data, and a concomitant shift in the Pt4f peaks from 71.4 eV (immediately following Pt deposition) to 72.4 eV (after CO exposure), see Figure S5.

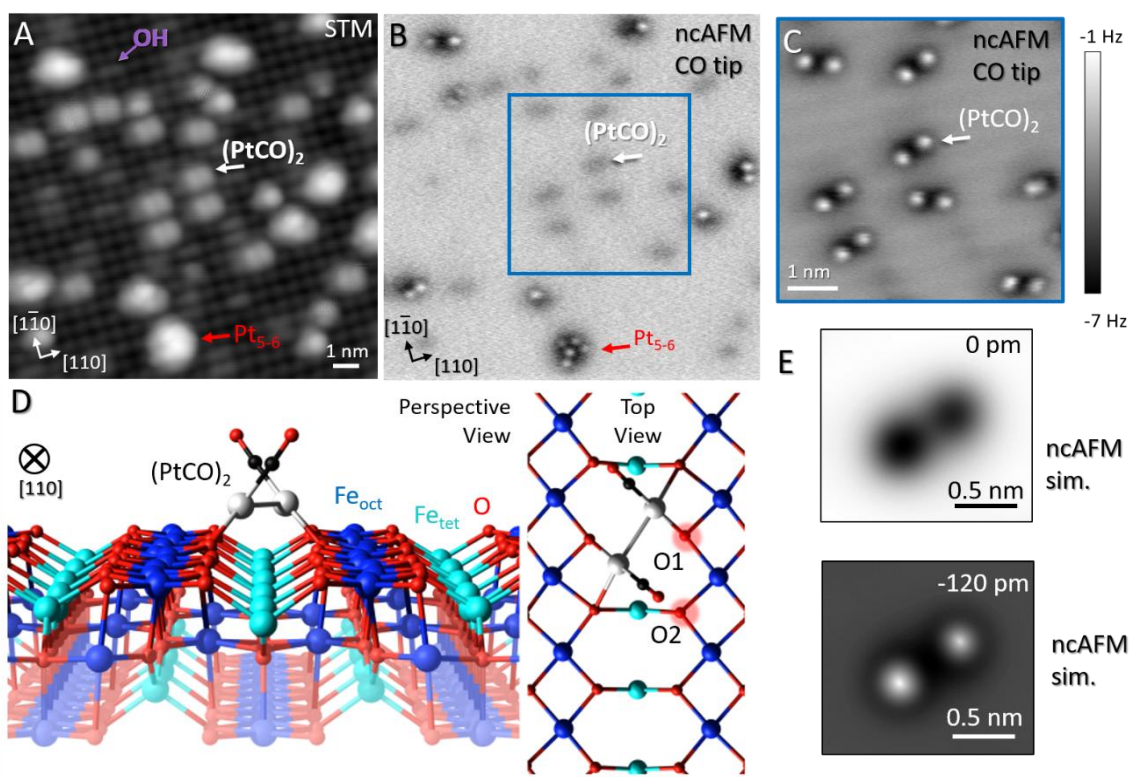

**Figure 1: Imaging the Pt/Fe₃O₄(001) system following CO-induced sintering. A**. STM image ($9 \times 9$ nm$^2$, $V_{sample}$ = +1.0 V, $I_{tunnel}$ = 2 pA, $T_{sample}$ = 77 K) obtained following exposure of a 0.2 ML ($2.84 \times 10^{13}$ Pt/cm$^2$) Pt/Fe₃O₄(001) model catalyst to CO at room temperature. CO-induced sintering leads to agglomeration of PtCO into clusters of various sizes, the majority of which are (PtCO)₂ dimers (white arrow). The red arrow highlights a cluster containing 5-6 Pt atoms. A surface hydroxyl group is also indicated in purple. **B.** Constant-height ncAFM image from the same sample area, taken with a CO-functionalized tip. CO molecules at larger Pt clusters are imaged as bright spots in repulsive interaction regime. The smaller (PtCO)₂ features are still in the regime of attractive forces, and appear dark. **C.** Constant-height ncAFM image ($4 \times 4$ nm$^2$) of the (PtCO)₂ dimers acquired at a closer tip-sample distance (≈120 pm closer compared to panel B). The CO molecules of each dimer are resolved, and appear rotated with respect to the underlying Fe rows of the Fe₃O₄(001) support, which appear dark. **D.** Perspective and top view of the (PtCO)₂ dimer on Fe₃O₄(001) as determined by DFT+U calculations. **E. n**cAFM simulations based on the structure shown in D, with different CO-surface separations.

## Noncontact Atomic Force Microscopy Imaging of Pt/Fe₃O₄(001).

In recent years, ncAFM has emerged as a tool to image surfaces and adsorbates with unprecedented resolution (*40-42*). Figures 1B shows ncAFM images of the same sample area as shown in Fig. 1A, obtained using a CO-terminated tip in constant-height mode. CO molecules adsorbed on the Pt₅₋₆ cluster are imaged as bright protrusions because the tip-sample distance is shorter to reach the repulsive regime of the interaction potential. The repulsion appears to be electrostatic, and linked to the opposing dipole moment of the CO molecules on the tip and the Pt cluster. At this tip height, the smaller $Pt_2$ species are imaged as faint dark protrusions, indicating the tip-sample distance is still in the regime of attractive interaction. When the tip is brought closer (panel C), two bright protrusions appear above each $Pt_2$ dimer. The separation of the protrusions is 600 pm, and the axis is rotated slightly with respect to the Fe rows of the support, which are imaged faintly dark at this tip-sample distance. Occasionally, the axis of a particular species flips between the two symmetric configurations during measurement. Otherwise, no mobility is observed at 78 K.

The bright protrusions observed in ncAFM measurements can be explained using the minimum energy configuration for a $(PtCO)_2$ species calculated by DFT+U (Figure 1D). Each Pt atom is bound to two surface oxygen atoms on neighbouring rows of the support structure, leading to the rotation of the Pt-Pt bond axis away from the [110] direction. The adsorbed CO molecules lean away from each other, and towards the opposite row of surface oxygen atoms. The predicted distance between the O atoms of the CO molecules is 520 pm, which is 80 pm less than the separation measured by AFM. This discrepancy likely arises from lateral bending of the CO molecules both at the tip and at the surface (*43, 44*).

**Reactivity Measurements Using Isotopically Labelled Temperature Programmed Desorption (TPD).** To investigate the reactivity of the $Pt/Fe_3O_4$ system we performed TPD experiments. First, the $Fe_3O_4$ single crystal sample was heated in $1\times10^{-6}$ mbar $^{18}O$ at 900 K for 3 hours, leading to a surface (mostly) isotopically labelled by $^{18}O$ (see low-energy ion scattering data in Fig. 2B). It has been shown (*45*) that Fe diffuses from the bulk to the surface under such conditions and reacts with $O_2$, leading to the growth of many layers of pristine $Fe_3O_4(001)$. Subsequently, 0.5 ML $Pt_1$ was deposited on a freshly prepared, isotopically labelled surface and exposed to $10^{-6}$ mbar.s $^{13}C^{16}O$ using an effusive molecular beam source (*46*) at room temperature. This creates an initial state similar to that shown in Fig. 1. $^{13}CO$ was utilized to easily differentiate reactant molecules from those of the residual gas, and to achieve the best possible signal/noise ratio. In the TPD experiment, the sample is heated from room temperature with a linear ramp of 1 $Ks^{-1}$, and the desorbing molecules are detected using a mass spectrometer in a line-of-sight geometry. During the TPD experiment no additional CO or $O_2$ is supplied. The first time this experiment is performed (black data in Fig. 2C), a steadily increasing desorption of m/e = 29 ($^{13}C^{16}O$) is observed between 300 K and 450 K, followed by a clear peak at 520-530 K. A similar peak is observed for $CO_2$ in both the m/e = 47 and m/e = 45 channels, corresponding to $^{13}C^{16}O^{18}O$ and $^{13}C^{16}O^{16}O$, respectively. The data shown in Fig. 2C (filled grey curve) is the sum of both contributions, but we note that ≈70% of the signal is of the $^{18}O$ labelled variety. These data suggest that most $CO_2$ is formed by extraction of isotopically labelled $^{18}O$ from the metal oxide support and that CO and $CO_2$ molecules most likely emerge from a common process on the surface at 520-530 K. The signal from $^{13}C^{16}O^{16}O$ arises because $^{16}O/^{18}O$ exchange occurs between surface and bulk during the TPD ramp and not the Boudouard reaction. This is clear because no C is detectable on the surface by XPS after the TPD experiment (see Figure S5).

To estimate the amount of $CO_2$ produced in the experiment, we calibrated the $CO_2$ peak area against a saturated monolayer of physisorbed $CO_2$, which has a known density of $5.68 \times 10^{14}$ cm$^{-2}$ on $Fe_3O_4(001)$ (*46*). This suggests that ≈$1.2 \times 10^{13}$ $CO_2$ molecules per cm$^2$ are formed, which is the same order as the Pt dimer coverage observed in STM after the first CO exposure (Fig. 2A).

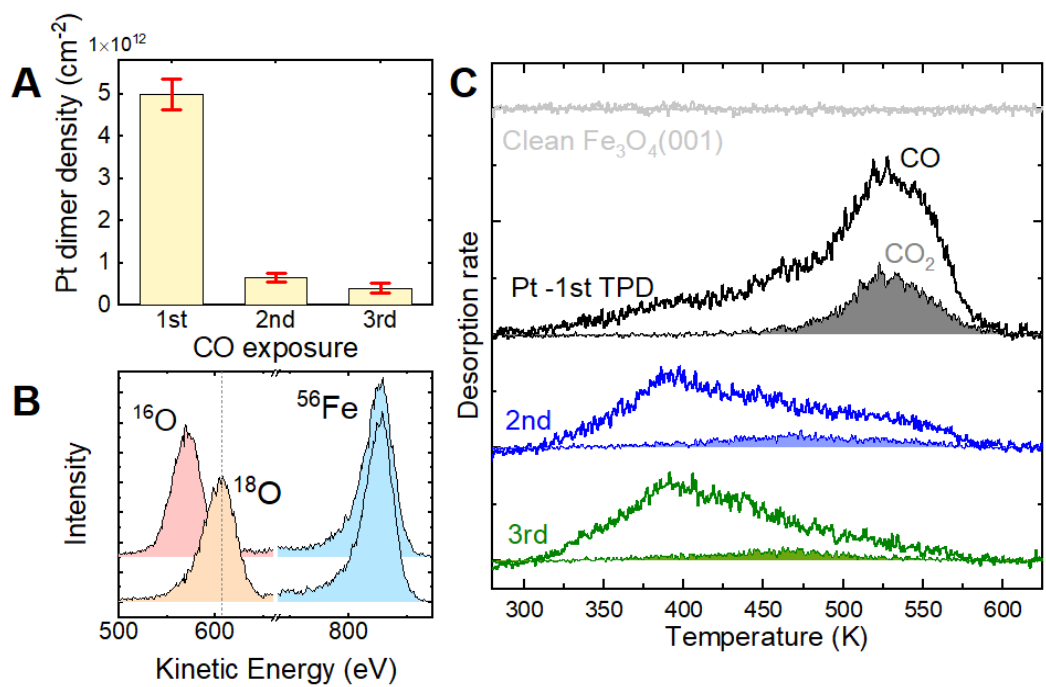

**Figure 2: Quantifying the reactivity of a Pt/Fe₃O₄(001) model catalyst. A.** Bar graph showing how the density of (PtCO)₂ dimers present on the surface changes prior to the 1$^{st}$, 2$^{nd}$ and 3$^{rd}$ TPD experiment. The densities were obtained from separate STM experiments for a Pt₁ coverage of 0.4 ML. **B.** Low-energy ion scattering data

(1 keV He$^+$) showing the isotopic composition of surface oxygen on the pristine surface (pink) and after heating the Fe$_3$O$_4$(001) sample at 900 K in $^{18}$O for 3 hours (orange). The $^{56}$Fe peak (blue) is unaffected by the procedure, as is the O:Fe ratio. **C.** Temperature programmed desorption data obtained from a 0.5 ML Pt/Fe$_3$$^{18}$O$_4$(001) sample following exposure to 1×10$^6$ mbar.s $^{13}$CO. A desorption peak for CO$_2$ at 520-530 K is observed in the first temperature excursion, in which (PtCO)$_2$ dimers were present in the initial state. Subsequent TPD cycles exhibit a $^{13}$CO desorption peak around 400 K, consistent with desorption of CO from Pt nanoparticles.

Somewhat surprisingly, if the sample is cooled to room temperature following the TPD experiment and then re-exposed to 10$^{-6}$ mbar.s CO, the desorption peak at 525 K vanishes from the CO and CO$_2$ spectra. Instead, a broad m/e = 29 $^{13}$CO signal is observed over the range of 300 – 550 K. A very low m/e = 47 $^{13}$C$^{16}$O$^{18}$O desorption signal peaks at ≈ 450-500 K. Further repetitions yield almost identical behaviour. Therefore, only the sample obtained following the initial sintering of the Pt$_1$ adatoms produces significant amounts of CO$_2$ at 520-530 K. To ascertain why, we imaged the surface using STM over a series of experiments mimicking the TPD, and counted the various Pt-containing species present at each step. For an initial coverage of 0.4 ML Pt$_1$ adatoms (5.68 ×10$^{13}$ Pt$_1$ cm$^{-2}$), the most common species on the surface following CO-induced sintering is the (PtCO)$_2$ dimer, with a coverage of 5±0.4 ×10$^{12}$ cm$^{-2}$. The (PtCO)$_2$ density falls sharply to (5±1) ×10$^{11}$ cm$^{-2}$ after heating to 510 K, and the resulting surface comprises a mixture of Pt$_1$ adatoms and larger clusters (see Figure S6). When this surface is exposed to CO, many of the Pt$_1$-CO species that are formed are captured by larger clusters, and the (PtCO)$_2$ density increases to (6±1) ×10$^{11}$ cm$^{-2}$. The broad CO desorption feature with a maximum at 400 K observed in the second TPD experiment is likely linked to these larger Pt clusters, which do not re-disperse on heating. The system does not change significantly with further CO exposure, so the third TPD resembles the second. To be sure that the surface did not evolve further between room temperature and the reaction temperature we heated the (PtCO)$_2$/Fe$_3$O$_4$(001) system to 550 K, and then imaged it at room temperature with STM (Fig. S6). This show that (PtCO)$_2$ dimers remain present in a similar density to found after sintering at room temperature. Based on these experiments, we conclude that the production of CO$_2$ during the first TPD is correlated with the presence of the (PtCO)$_2$ species in a quantitative manner.

**Density Functional Theory Calculations to Determine the Reaction Pathway.** To understand how (PtCO)$_2$ catalyzes CO oxidation, we performed DFT calculations. Based on



the excellent agreement between experimental and simulated ncAFM images shown in Figs. 1C and 1E, we are confident to have obtained the correct minimum-energy configuration for a $(PtCO)_2$ dimer. Since TPD shows that CO oxidation clearly occurs through extraction of lattice oxygen, we determine the minimum energy path (MEP) leading to $CO_2$ utilizing oxygen from the support lattice. To assess the validity of the MEP, and to distinguish between mechanisms involving different entropic effects, we employed a micro-kinetic model, which bridges the 0 K DFT results and finite-temperature experiments. Ultimately, the kinetic model yields a predicted desorption temperature that can be compared quantitatively to the TPD experiments. Based on our experience studying CO TPD from a variety of adatoms supported on $Fe_3O_4(001)$[19], as well as benchmarking done specifically for this study (see Supplementary Information), we expect *quantitative agreement* between experiment and theory within 30 K (DFT values too high most likely due to a slight over-binding at DFT level), and this serves as a stringent criterion with which to judge the different reaction mechanisms.

Inspecting the $(PtCO)_2$ geometry shown in Fig. 1D, we notice that the adsorbed CO molecules each lean towards an "O2" atom. Calculations based on the commonly-used nudged elastic band (NEB) method [47] find no barrier to form OC-O (see Movie Path-C), but the final state following $CO_2$ desorption is sufficiently unfavorable that the kinetic model predicts that $CO_2$ would not desorb until 700 K in a TPD experiment. The ≈175 K difference between theory and experiment suggests that another, more favorable pathway must exist. Another possibility is that the $(PtCO)_2$ splits into two adatoms, and that the PtCO species react independently with the surface. The barrier to split the stable $(PtCO)_2$ is high (2 eV), however, and the process should not occur until 680 K.

Since the observed $CO_2$ evolution cannot be explained using the minimum-energy $(PtCO)_2$ configuration, we next consider the possibility that the dimer can adopt a non-equilibrium morphology at elevated temperature. This isomerization, sometimes referred to as fluxionality[48], has been demonstrated to lower the overall energy required for a reaction to proceed for sub-nano clusters [8, 48]. We constructed an alternative $(PtCO)_2$ configuration (see inset B in Fig. 3 and Supplementary Movie path-B) based on two sites previously observed to be stable for isolated Pt adatoms on $Fe_3O_4(001)$ [39]. The first, with one Pt twofold coordinated to surface oxygen midway between the surface Fe rows, is the site



typically observed for single metal adatoms on $Fe_3O_4$(001). The second Pt is twofold coordinated to surface oxygen along the direction of the surface Fe rows, and resembles a metastable configuration directly observed for $Pt_1$/$Fe_3O_4$(001) (*39*). In this configuration, the CO molecules adopt positions such that they are not in close proximity, and the total energy is 0.5 eV higher than the ground state (configuration B in Fig. 3). The barrier to reach this configuration was calculated using the NEB approach, and found to be at most 1.5 eV. In reality the barrier may be lower because we considered sequential motions for the Pt atoms and the real process will have concerted movement. In any case, the 1.5 eV barrier already means that the metastable cluster is accessible at 550 K, which is, within the DFT error bars mentioned above, compatible with experiment. Nevertheless, we find that extracting the oxygen atoms within reach of the CO remains energetically prohibitive, this time because it is difficult to extract surface oxygen atoms to which the Pt is bound. However, direct CO desorption can occur from this configuration at 570 K (black pathway in Fig. 3, see Movie path-A2), culminating in configuration F2.

In what follows, we describe the reaction pathway that, according to our analysis, leads to CO oxidation in the $(PtCO)_2$/$Fe_3O_4$(001) system (blue path in Fig. 3 and Movie Path-A). This process utilizes the $(PtCO)_2$ configuration described above, together with a metastable configuration of the $Fe_3O_4$(001) support. Starting from the metastable $(PtCO)_2$ configuration Fig. 3B, one Pt atom approaches a surface Fe site, displacing the Fe into a subsurface interstitial site with tetrahedral coordination. This phenomenon, which has been observed for a variety of metal adatoms on this surface including Ir (*49*) and Rh (*27*), induces a cascade of Fe diffusion events between neighboring octahedral and tetrahedral sites in the subsurface, leading to occupation of both of the third layer Fe vacancies characteristic of the $Fe_3O_4$(001) surface reconstruction (*36*). Configuration (C) represents the transition state, which lies 1 eV above the initial state, i.e., it can easily happen at 550 K. The next relevant step (E) is for the CO molecule to form a bond to a surface oxygen atom on the far side of the Fe row, which is a barrier-less process. The formation of a $CO_2$ molecule costs 1.2 eV, while desorption costs a further 0.4 eV. This includes the creation of a surface oxygen vacancy, the entropic gain through desorption at the reaction temperature, and the cost of accommodating the $Pt_2CO$ on the surface after the reaction. The cost of the final state is minimized by accommodating one Pt atom in a substitutional site within the local $Fe_3O_4$



surface structure (configuration F in Fig. 3), while the other occupies a bulk-continuation site twofold coordinated to surface oxygen atoms. Thus, the entire process to desorb $CO_2$ requires about 1.6 eV, which corresponds to a TPD peak maximum for $CO_2$ at 545 K. This is approximately 20-30 K higher than observed experimentally, i.e., within the expected error bars (see SI).

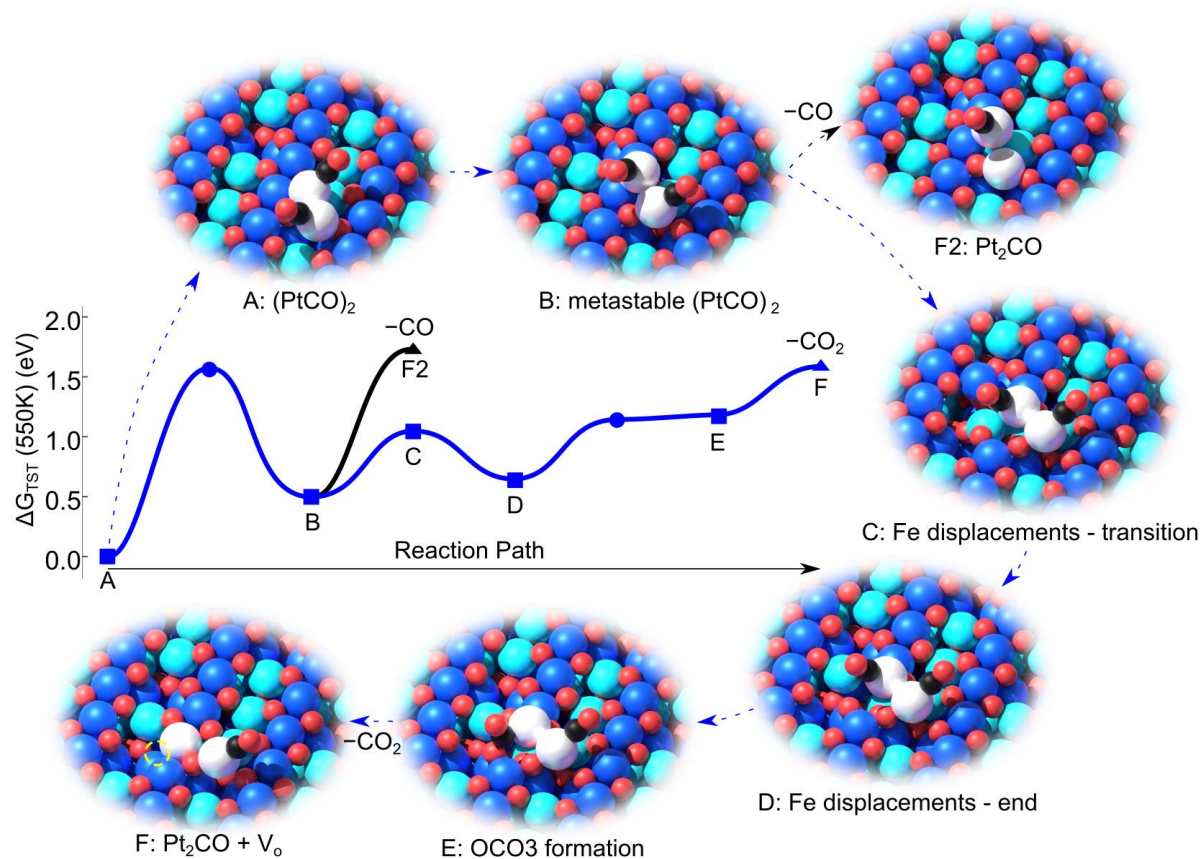

**Figure 3: Proposed reaction scheme for the (PtCO)₂ species determined by DFT calculations (see also Movie Path-A).** In the schematics, the $Fe_{oct}$ and $Fe_{tet}$ of the $Fe_3O_4$(001) support are dark blue and cyan, respectively. O atoms are red, Pt are white, and the C and O in CO are black and red, respectively. **A.** The stable (PtCO)₂ dimer (see Fig. 1D) is the reference for 0 energy. **B.** Alternative (PtCO)₂ configuration in which one Pt atom moves to become 2-fold coordinated to surface undercoordinated O across the surface rows (O1 in Fig. 1D). The second Pt atom remains coordinated to two neighbouring surface O atoms along the [110] direction (O1 and O2 in Fig. 1D). The CO molecules reorient but remain tilted away from each other. **C-D.** The presence of the metastable (PtCO)₂ pushes a surface $Fe_{oct}$ into an interstitial site with tetrahedral coordination (thus, it becomes cyan in the model). This goes hand in hand with a cascade of Fe diffusion events in the subsurface culminating in $Fe_{int}$ occupying a third layer iron vacancy. In (D), one Pt substitutes the missing surface $Fe_{oct}$. **E.** CO binds to the undercoordinated surface O in a barrier-less process. **F.** $CO_2$ desorption leaves a surface oxygen vacancy ($V_O$, broken circle) and one Pt occupying an $Fe_{oct}$ site. **F2.** Desorption of CO from the metastable Pt dimer is competitive with $CO_2$ formation.



The desorption of the $CO_2$ molecule leaves behind a metastable $Pt_2CO$ species. Desorption of the CO is energetically unfavorable (CO adsorption energy at the $Pt_2CO$ is -3.37 eV), and it is much easier to break the $Pt_2CO$ into a twofold coordinated $Pt_1$ adatom and a PtCO (see Figure S4). Since PtCO are already observed to diffuse readily at 300 K in STM (*39*), they will be highly mobile at the reaction temperature. If they meet a second PtCO, a $(PtCO)_2$ species will be formed and the same reaction repeats. Alternatively, mobile PtCO may first encounter existing Pt clusters, leading to coalescence and the immediate liberation of the adsorbed CO (CO is seen to desorb around 400 K from Pt clusters on the basis of the repeat TPD experiments in Fig. 2C.) This explains why desorption of CO and $CO_2$ is simultaneous in TPD, and why a mixture of $Pt_1$ adatoms and larger Pt clusters is observed in post-TPD STM experiments (*39*). The full energetics of the pathway after $CO_2$ desorption are shown Fig. S4.

Finally, we address the $CO_2$:CO ratio. While the reaction described in Fig. 3A-F would be expected to yield a 1:1 $CO_2$:CO ratio, the alternative branch shown in black leads exclusively to CO. The kinetic model suggests that the direct CO desorption should yield a TPD peak at approximately 575 K, which is some 30 K higher than the $CO_2$ pathway. This is consistent with the weak high-temperature shoulder visible in the CO data in Fig. 2C, as well as the $CO_2$:CO ratio smaller than 1:1 in the TPD experiments in the 500–550 K TPD peak.

## Discussion

Based on the experimental and theoretical evidence presented above, we conclude that $Pt_2$ dimers catalyse the initial CO oxidation activity in what was nominally a $Pt_1/Fe_3O_4(001)$ model system (in the absence of CO). The main reason for this is that CO-induced sintering is facile even at room temperature,(*39*) and no isolated $Pt_1$ adatoms remain at the reaction temperature. Our STM/AFM experiments image the $(PtCO)_2$ dimer initial state with exquisite resolution, allowing us to have a high degree of confidence in the structure predicted by DFT calculations. Isotopically labelled TPD data clearly demonstrate that lattice oxygen is extracted to form $CO_2$, and STM shows the final state after the TPD is a mixture of single Pt adatoms and clusters. Together with the extensive benchmarking of theory (see SI), these results provide a stringent test of the mechanism proposed computationally. Many simpler, seemingly plausible pathways (see details in the SI) were rejected on the basis that



$CO_2$ would not evolve at the correct temperature, and that the final state would not match that observed in experiment.

The most unexpected aspect of the proposed mechanism is that the reaction occurs when both the $(PtCO)_2$ species and $Fe_3O_4$ support enter a metastable configuration. Metastable cluster geometries are increasingly invoked to explain the reactivity of sub-nano clusters[8, 48], and it is necessary here because extracting the O atoms within reach of the equilibrium $(PtCO)_2$ structure is energetically unfeasible. Our work suggests that the support cannot be treated as static either, and that concerted rearrangements of the cluster and support must be taken into account. The idea to allow for subsurface Fe mobility originally arose from prior experimental observations, as several other metals have been shown to displace surface Fe into the subsurface layers on $Fe_3O_4(001)$ even at room temperature (*50*). Of course, considering a metastable support in addition to a metastable cluster widens the possible reaction pathways substantially, and it would have been extremely difficult to arrive at the final mechanism by theory alone. We conclude that combining theory with high quality, unambiguous experimental data is crucial to guide the computations through the vast landscape of possibilities.

In Fig. S2 we show alternative pathways for a PtCO species adsorbed on $Fe_3O_4(001)$, to better understand whether the observed CO oxidation activity could, in principle, have emerged from such species. Interestingly, we find that lattice O extraction leading to $CO_2$ can occur at an isolated $Pt_1$ site at ≈450 K, and that the process is competitive with CO desorption. Thus, in the absence of the experiments, it would have been possible to erroneously conclude that the $Pt_1/Fe_3O_4(001)$ is an active SAC system. In agreement with experiment, however, we find that the barrier for PtCO diffusion is significantly lower than that of reaction, meaning that agglomeration at the temperature required for lattice O extraction would be extremely rapid, even at low coverage. Thus, it is important that diffusion barriers of intermediate states should be routinely calculated in SAC screening studies, particularly when CO is involved.

Recently, the Sykes group (*32*) studied $Pt_1$ on an ultrathin copper-oxide film grown on Cu(111) using a similar approach. In agreement with our results, this UHV-based study concluded that the oxide-supported $Pt_1$ adatoms have an almost neutral charge state (*32, 39*), whereas most reports of $Pt_1$ atoms adsorbed on metal oxide supports in the SAC



literature conclude a $Pt^{2+}$ or $Pt^{4+}$ state. This assignment is usually based on the observation of a 40-50 $cm^{-1}$ blue shift of adsorbed CO in infrared absorption spectroscopy (IRAS) studies (*22, 23*), but this assignment is not without controversy (*25*). It is well known that the oxidation state of isolated adatoms can be changed by CO (*51*), and we observe a positive shift in the Pt 4f binding energy in XPS when CO is adsorbed (see Fig. S5 and refs. (*19, 39*)). This is in line with our calculations, which suggest that CO adsorption decreases the Pt Bader charge by 0.3 $e^-$. Ultimately, $CO_2$ evolution was observed at lower temperature on the CuO film because it is easier to extract O from the copper oxide than $Fe_3O_4$. Utilizing a more reducible oxide support allows $O_{lattice}$ extraction to proceed at lower temperature, with the added benefit of less thermal sintering.

Finally, since utilizing $Pt_1/Fe_3O_4(001)$ as a single-atom catalyst is clearly hampered by PtCO diffusion, it is tempting to consider whether other metals might fare better. We have recently shown that a model $Ir_1/Fe_3O_4(001)$ is stable against CO-induced sintering, primarily because the formation of carbonyl/dicarbonyl species creates a stable pseudo-square-planar environment for the cation (*49*). CO binds significantly more strongly to $Ir_1$ than $Pt_1$ on $Fe_3O_4$, which would usually be seen as problematic from the point of view of CO poisoning. However, strong CO binding is actually advantageous for a MvK mechanism as it ensures that CO remains at the surface at temperatures where facile extraction of lattice oxygen can occur. This is consistent with results obtained on the $Ir_1/FeO_x$ system (*13*), where better water-gas shift (WGS) reaction performance compared to $Pt_1/FeO_x$ was linked to a MvK mechanism.

In summary, we have shown that $Pt_2$ dimers, formed dynamically on a $Pt/Fe_3O_4$ model catalyst facilitate the extraction of oxygen from the support lattice at 525 K. The energy required is minimized when both cluster and support adopt non-equilibrium configurations, highlighting the need to consider dynamic restructuring for reactions occurring at elevated temperatures. Ultimately, our work is a clear demonstration that metastable active species can form upon exposure to gases, and that the addition of just one atom can make a big difference to a "single-atom" catalyst.



**Methods**

STM and ncAFM experiments in Fig. 1 were performed at T=78 K using a commercial Omicron LT-STM using a qPlus sensor (k = 1800 N/m, $f_0$ = 31000 Hz, Q ≈ 10000) with a separate wire for the tunneling current, and a differential cryogenic preamplifier[52]. Electrochemically etched W tips were glued to the tuning fork and cleaned in-situ by field emission and self-sputtering in $1 \times 10^{-6}$ mbar argon [53]. The tip was functionalized by picking CO from atop a Pt-CO cluster. Such functionalization is stable enough to allow imaging at 78 K. TPD and XPS experiments were conducted in a different vacuum system with a base pressure of $\approx 5 \times 10^{-11}$ mbar [46]. In both setups, the $Fe_3O_4$(001) single crystal (SurfaceNet GmbH) was prepared by cycles of room-temperature 1 keV $Ne^+$ sputtering followed by annealing at 650 °C. Every other annealing cycle was conducted in an $O_2$ partial pressure of $1 \times 10^{-6}$ mbar. Pt was evaporated directly onto the sample surface using a Focus EFM3 evaporator, with the flux determined by a water cooled quartz crystal microbalance. For the TPD/XPS experiments, CO was dosed using a calibrated molecular beam source, which is described in detail in ref. (46), along with the rest of the experimental TPD setup.

The Vienna *ab initio* Simulation Package (VASP) (*54, 55*) was used for all calculations. The Projector Augmented Wave (PAW) (*56, 57*) method describes the near-core regions; and the plane wave basis set cut-off energy was set to 550 eV. A Γ-centred **k**-mesh of 5 × 5 × 5 was used for the bulk, $Fd\overline{3}m$, a= 8.396Å, experimental lattice magnetic cell, adjusted to 1 × 1 × 1 for (001) surface calculations. The optB88-DF van der Waals functional(*58, 59*) was used with an effective on-site Coulomb repulsion term $U_{eff}$ = 3.61 eV (*60, 61*) to accurately model the $Fe_3O_4$. Calculations were performed on an asymmetric slab with 13 planes (5 fixed and 2 relaxed $Fe_{oct}O_2$ layers) and 14 Å vacuum. Convergence is achieved when forces acting on ions become smaller than 0.02 eV/Å. To avoid interaction between adsorbates, and to accurately model the experimental coverages, the $Fe_3O_4$(001)-(2×2) supercell contained 380 atoms (i.e. four times the (√2×√2)R45° reconstructed cell was used). This computationally expensive setup is required for two reasons: First, a (2×2) supercell allows an accurate representation of the experimental Pt coverage. (Calculations performed on a (1×1) cell yielded generally lower adsorption energies hinting at a repulsive interaction.) Second, the supercell provides two adsorption sites for $Pt_1$ adatoms, which allows us to perform nudged



elastic band (NEB) calculations (*47*) to determine the barriers related to (PtCO)$_2$ formation, diffusion, and splitting.

The NEB calculations have been performed at the PBE+D2 level, because our system tends to get stuck in local minima when using optB88-DF. More precisely, the climbing image method (CI-NEB) was used to determine the saddle points. Long diffusion processes were split into multiple smaller NEB calculations, each holding 3 intermediate configurations. Intermediates of reactions such as different positions along diffusion paths or OCO intermediates were calculated utilizing optB88-DF, essentially taking advantage of its tendency to get trapped in local minima. Transition states obtained using PBE+D2 CI-NEB calculations were recalculated using optB88-DF with the atomic positions fixed to obtain the corresponding energies. The simulated AFM images shown in Fig.1E were obtained using the AFM simulation toolkit within the probe particle model (*44*).

The energies shown in Figure 3 are given in terms of Gibbs free energies at 550 K, with the simplifying assumption that on-surface movements have zero entropy change. The entropy change upon desorption is accounted for in the framework of the transition state theory (*62, 63*), which reduces the energy required to desorb molecules relative to the 0 K reaction pathway (all on-surface processes are identical to the 0 K reaction pathway). The ΔG values are then used together with the steady state approximation (*64*) to estimate the rates of different processes as a function of temperature (assuming a 1 Ks$^{-1}$ ramp, as in the experiment), which ultimately allows us to estimate the TPD peak temperature for each potential desorption event. Movies illustrating the processes involved are included in the Supporting Information. All calculated rates and deduced temperatures at which shown processes can occur are listed and discussed in the SI with more details.


### Acknowledgements

GSP, MM, and FK acknowledge funding from the European Research Council (ERC) under the European Union's Horizon 2020 research and innovation program Grant agreement No. 864628. JH, RB and ZJ were supported by the Austrian Science Fund (FWF, Y847-N20, START Prize). ZJ acknowledges a stipend from the TU Wien doctoral college TU-D. GSP, UD, MS and CF acknowledge funding from the FWF SFB TACO (F81).




**Author Contributions.**

MM performed and evaluated theoretical calculations under the supervision of CF. JH, ZJ, FK, MB, RB and MSetvin performed experiments and analysed data. MSchmid, UD, and GSP acquired funding, planned the research and drafted the manuscript.

**Competing Interests**

The authors declare that they have no competing interests.

**Data Availability**

All data needed to evaluate the conclusions in the paper are present in the paper and/or the Supplementary Materials.

**Supplementary Materials**

1- Kinetic model details

2- Additional computationally studied reaction paths

3- Experimental data showing Pt4f and C1S XPS data from $Pt_1/Fe_3O_4(001)$ before and after CO exposure, TPD data for different Pt coverages, and room temperature STM images acquired after heating the $(PtCO)_2/Fe_3O_4(001)$ system to 550 and 580 K

4- Supplementary Movie Captions





Diffusion_PtCO_path-across.mp4 and Diffusion_PtCO_path-along.mp4

# CO oxidation by Pt$_2$/Fe$_3$O$_4$: metastable dimer and support configurations facilitate lattice oxygen extraction

**Authors:** Matthias Meier[1,2†], Jan Hulva[1†], Zdenek Jakub[1], Florian Kraushofer[1], Mislav Bobić[1], Roland Bliem[1‡], Martin Setvin[1,3], Michael Schmid[1], Ulrike Diebold[1], Cesare Franchini[2,4] and Gareth S. Parkinson[1]

[1]Institute of Applied Physics, TU Wien, Vienna, Austria.

[2]Computational Materials Physics, University of Vienna, Vienna, Austria.

[3]Department of Surface and Plasma Science, Faculty of Mathematics and Physics, Charles University, Prague, Czech Republic

[4]Alma Mater Studiorum – Università di Bologna, Bologna, Italy.

*Correspondence to: parkinson@iap.tuwien.ac.at

†These authors contributed equally to this work

‡Current address: Advanced Research Center for Nanolithography, Science Park 106, 1098XG Amsterdam, The Netherlands

**List of contents:**

**1- Kinetic model details**
**2- Additional computationally studied reaction paths**
    1- Benchmarking DFT
    2- PtCO diffusion
    3- PtCO reactions
    4- $(PtCO)_2$ paths
    5- $Pt_2CO$ (post-$CO_2$ desorption) path
**3- Experimental data showing Pt4f and C1S XPS data from $Pt_1/Fe_3O_4(001)$ before and after CO exposure, TPD data for different Pt coverages, and room temperature STM images acquired after heating the $(PtCO)_2/Fe_3O_4(001)$ system to 550 and 580 K**
**4- Supplementary Movie Captions**

**List of figures:**

Fig. S1: Comparing CO desorption energies obtained from TPD experiments with DFT+U determined adsorption energies
Fig. S2: PtCO reactivity and $(PtCO)_2$ formation
Fig. S3: $(PtCO)_2$ alternatives
Fig. S4: $Pt_2CO$ post-$CO_2$-desorption path
Fig. S5: Pt4f and C1S XPS data from $Pt_1/Fe_3O_4(001)$ before and after CO exposure
Fig. S6: STM images of the $Pt/Fe_3O_4(001)$ system following CO exposure and heating to temperatures (A) 553 K and (B) 580 K.
Fig. S7: CO TPD from different coverages of Pt on $Fe_3O_4(001)$.

**List of animations:**
**Paths relative to Fig. S3, shown as short animations:**
path A.mp4
path A2.mp4
path B.mp4
path C.mp4
path D.mp4
path E.mp4
**Paths relative to Fig. S4 (post-$CO_2$), shown as a short animation:**
path F.mp4
**Paths relative to PtCO diffusion, shown as short animations:**
Diffusion_PtCO_path-across.mp4
Diffusion_PtCO_path-along.mp4

**1-Kinetic model details**

The kinetic model applied here is based on the previous work of Campbell et al. (63), as in our previous work (19) on CO desorption from model single-atom catalysts (SAC). There, we were concerned with direct barrier-less desorption processes only. In the current work, we need to compare these direct desorption events to complex, multi-step and reversible processes such as OCO formation (potentially being the rate-limiting step), $CO_2$ desorption, and diffusion processes (*e.g.* the split of $(PtCO)_2$ dimers into two adatoms). We do this within the steady state approximation.

The kinetic model consists of solving the well-known Polanyi-Wigner equation:

$$p \propto \frac{-d\theta}{dt}, \text{ where } \frac{-d\theta}{dt} = \theta k \text{ and } \frac{-d\theta}{dt} = \frac{-d\theta}{dT}\beta \qquad \text{eqn. S1}$$

with $p$ the partial pressure, $k$ the reaction rate constant, $\beta$ the temperature ramp in $K.s^{-1}$, and $\theta$ the coverage of the initial state (here shown for a first order barrier-less desorption process).

The rate constant is calculated with the usual Arrhenius equation:

$$k = \frac{k_B T}{h} exp\left(\frac{\Delta S_{TST}}{k_B}\right) exp\left(\frac{\Delta H_{TST}}{k_B T}\right) \qquad \text{eqn. S2}$$

The entropy term ($\Delta S_{TST}$, within transition state theory (TST)) for a barrier-less desorption process takes into account the change in entropy between the adsorbed phase and the entropy of the free molecule in the gas phase. For the gas phase, we used partition functions from the National Institute of Standards and Technology (NIST) database, and removed the translational contribution to S along the desorption path, *i.e.*, normal to the surface (63). The entropy of the adsorbed phase is taken to be 0, as we have previously shown that the contribution of CO vibrations to the entropy are very small with respect to the gas phase entropy (19). For a diffusion process, or any other process where both initial and transition states remain in the surface potential, we take $\Delta S_{TST}=0$. The enthalpy term ($\Delta H_{TST}$) is directly correlated to the calculated $\Delta E$ term at DFT 0 K-level, either using the barrier calculated from cl-NEB calculations, or the desorption energy. Additionally, we add the contributions to the enthalpy from the different partition functions if present (gas phase, translational and rotational terms).

The steady state approximation is used to treat multi-step processes, taking into account reversibility. Reaction rate constants and coverages are calculated iteratively as the temperature ramp (step in the model) is set equal to 1 $K.s^{-1}$, as in the TPD experiment. The kinetic model ultimately allows us to clearly determine the minimum energy path (MEP) at finite temperature starting from the initial configuration (*e.g.,* a $(PtCO)_2$ dimer, see Fig. 1D).

Experimentally observed coverages (0.5 ML of PtCO, 0.035 ML of $(PtCO)_2$) are applied to our model when simulating respective TPD spectra. All displayed energy diagrams are therefore taking into account entropic effects, which are calculated at 550 K (approximately the desorption temperature in experiment). The temperature of choice corresponds to the observed experimental temperature at which $CO_2$ is observed to desorb.

**2-Additional computationally studied reaction paths**

The aim of this section is to further support the credibility of our DFT-suggested mechanisms and paths, as well as underline the quantitative agreement with the experimental results (parts 2.1 and 2.2). Additionally, we show the hypothetical minimum energy path (MEP), for adatoms, in case dimers would not be formed (part 2.3). Finally, in the last part (2.4), we discuss alternatives to the main pathways mentioned in the main text.

<u>2.1- Benchmarking DFT</u>

While investigating this system, many reaction paths leading to $CO_2$ were found. In the majority of cases, the DFT-based kinetic model predicted temperatures that were too high with respect to the experimental observation of 520 K. (We show some of these alternatives in section 2.3 for discussion purposes.) But an important question is: how high is too high? Thus, it is important to estimate what constitutes a reasonable offset between the computationally determined temperatures and the experimentally. Here, we discuss a selection of results that allow us to estimate that a 30 K overestimation of the desorption temperature constitutes acceptable agreement.

The first such data, recently published (19) and reproduced here in Fig. S1, are a comparison of experimental CO TPD data for $Au_1$, $Ag_1$, $Cu_1$, $Ni_1$, $Rh_1$, and $Ir_1$, species supported on $Fe_3O_4(001)$. The computational details are the same as utilized here. As can be seen in Figure S1, it was found that the computational predictions overbind CO at the model SAC sites by approx. 0.25 eV, which corresponds to a simulated TPD peak roughly 30 K too high in temperature output by the kinetic model.

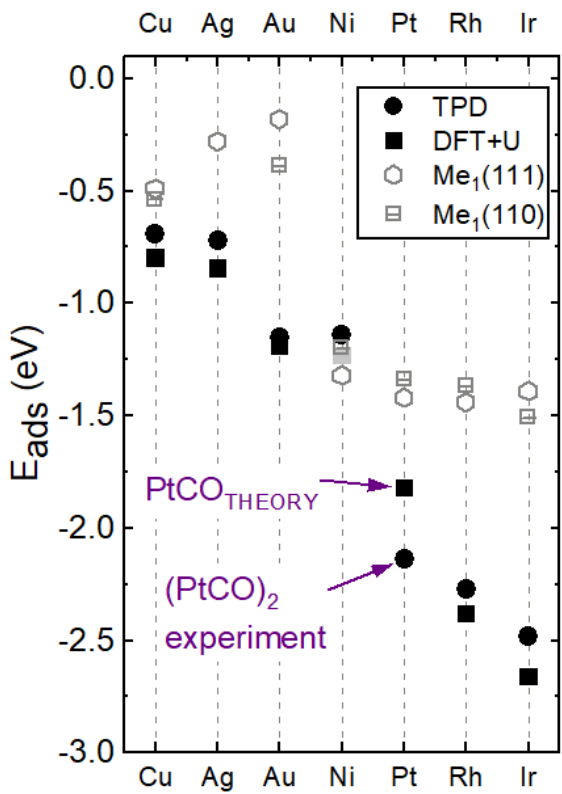

**Figure S1. Expected agreement between experimental and theoretical desorption temperatures.** Comparison of CO binding energy extracted from TPD data in Ref. 19 with the results of DFT+U calculations. For Cu, Ag, Ni, Rh and Ir, the computations predict a binding energy slightly (0.25 eV) in excess of the experimental result. Note that the trend is reversed for Pt. This is because the Pt adatoms sinter and form $(PtCO)_2$ dimers, which bind CO more strongly than Pt adatoms.

## 2.2- PtCO diffusion

The diffusion of PtCO and $Pt_1$ on $Fe_3O_4(001)$ were investigated computationally. The diffusion path of PtCO is shown in two short animations (supplementary movies Diffusion_PtCO_path-across and –along. Diffusion across the Fe rows, or along the Fe rows can occur with a similar barrier (Diffusion along Fe rows: 1.18 eV, diffusion across Fe rows: 1.21 eV) and thus occur at similar temperatures. This is in agreement with STM data, where no preferential diffusion direction was observed (39). However, since we are only able to observe infrequent diffusion events in room temperature STM movies with several minutes between consecutive scans/images, the rate for this process must be very low. Assuming a pre-exponential factor of $10^{13} \ s^{-1}$, the range of temperatures at which the PtCO diffusion rate is small but non-zero corresponds to a temperature of 340–350 K, which is 40–50 K above room temperature. This suggests that our calculated temperatures for diffusion processes are similarly overestimated than those calculated for direct desorption processes, and that a reaction pathway featuring a positive offset of 30–40 K could indicate the correct reaction pathway for $CO_2$ production. The diffusion of $Pt_1$ adatoms was studied using a similar approach and found to occur only at 880 K. This is why $Pt_1$ species are never observed to diffuse on $Fe_3O_4(001)$ in room-temperature STM movies (39).

## 2.3- PtCO reaction paths

In room-temperature STM movies, PtCO species are observed to diffuse and form $(PtCO)_2$ (39). In the left panel of figure S2, we show how this process happens from a computational perspective. The results indicate that the rate-limiting step is the diffusion of an isolated PtCO species, and there is no additional energy cost involved in the formation of $(PtCO)_2$ dimers.

In the right panel, we show the best alternative pathways for PtCO at room temperature that would lead to CO desorption and $CO_2$ formation/desorption. The dashed line corresponds to the maximum barrier calculated in the left panel. We see that diffusion is preferred to CO oxidation reaction at 325 K, but that the difference is within the error of our calculations. Given that no $CO_2$ (and very little CO) evolution is observed at temperatures below 525 K in the TPD experiment, we conclude that the trend is correct, and that the reaction to form $CO_2$ does not occur before diffusion leads to the formation of $(PtCO)_2$ species.

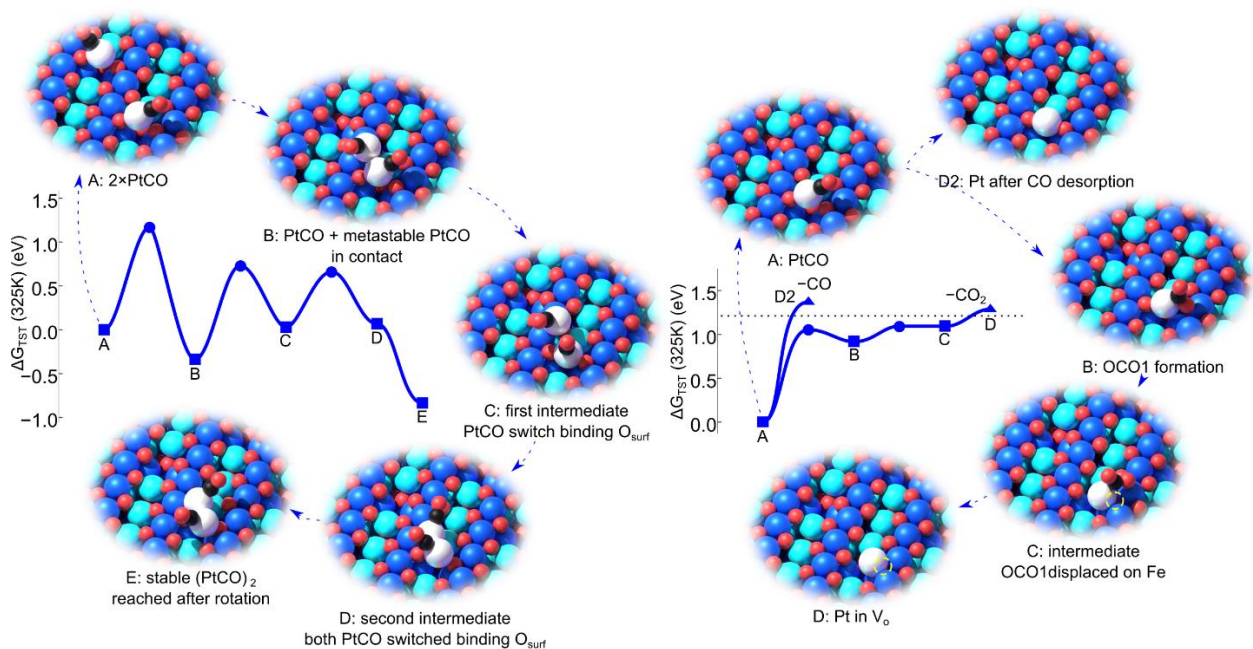

**Figure S2: (Left) Pathway for formation of a (PtCO)₂ from two PtCO species A-B.**
Two initially separate PtCO species exist in neighboring unit cells. In both, the Pt atoms are 2-fold coordinated to surface O1 atoms (see Fig. 1D in the main text for O atom designations). One of the PtCO becomes mobile and diffuses across the Fe row and forms a Pt-Pt bond with the stationary PtCO The diffusion path for a single species is also shown in Movie "Diffusion_PtCO_path-across". **C-E.**
The (PtCO)₂ species formed in B is not in its minimum-energy configuration, and thus rearranges such that each of the Pt atoms is ultimately bound to one O1 and one O2 (see Fig. 1D for O atom designations). This involves the breaking and formation of bonds, each of which has its own associated barrier. Nevertheless, each barrier is smaller than the initial diffusion barrier, and the final state E will thus be reached spontaneously. This final state is significantly downhill from the two separated PtCO species, making the (PtCO)₂ highly stable. **(right) Alternative reaction pathways for PtCO. A-D.**
Reaction of PtCO to form CO₂ via O_lattice extraction. The adsorbed CO first moves to form an OCO intermediate via a small barrier (B). Then the Pt moves towards the space vacated by O1, as this atom is lifted away from the surface (C). The CO₂ then desorbs, leaving a Pt atom occupying an oxygen vacancy (D). **(D2)** Path D2 represents the pathway for desorption of CO from the PtCO species. The dashed line is the barrier calculated for PtCO diffusion. In this figure the squares represent calculated converged configurations, circles represent cl-NEB climbing image saddle points, and triangle-ups represent desorption processes.

## 4- (PtCO)₂ paths

In Figure S3, we describe a few alternative processes to the best CO₂ forming path described in the main text, which is shown as the blue curve (path A). The alternative paths (red curves, paths B, C, D and E) are conceptually simpler, but the predicted temperatures from the kinetic model are not in quantitative disagreement with the experimentally observed TPD temperature. In some cases, these paths also lead to a final state that would be in disagreement with post-TPD STM data, i.e., a mixture of isolated Pt₁ adatoms and larger clusters.

Figure S3 shows a selection of alternatives.

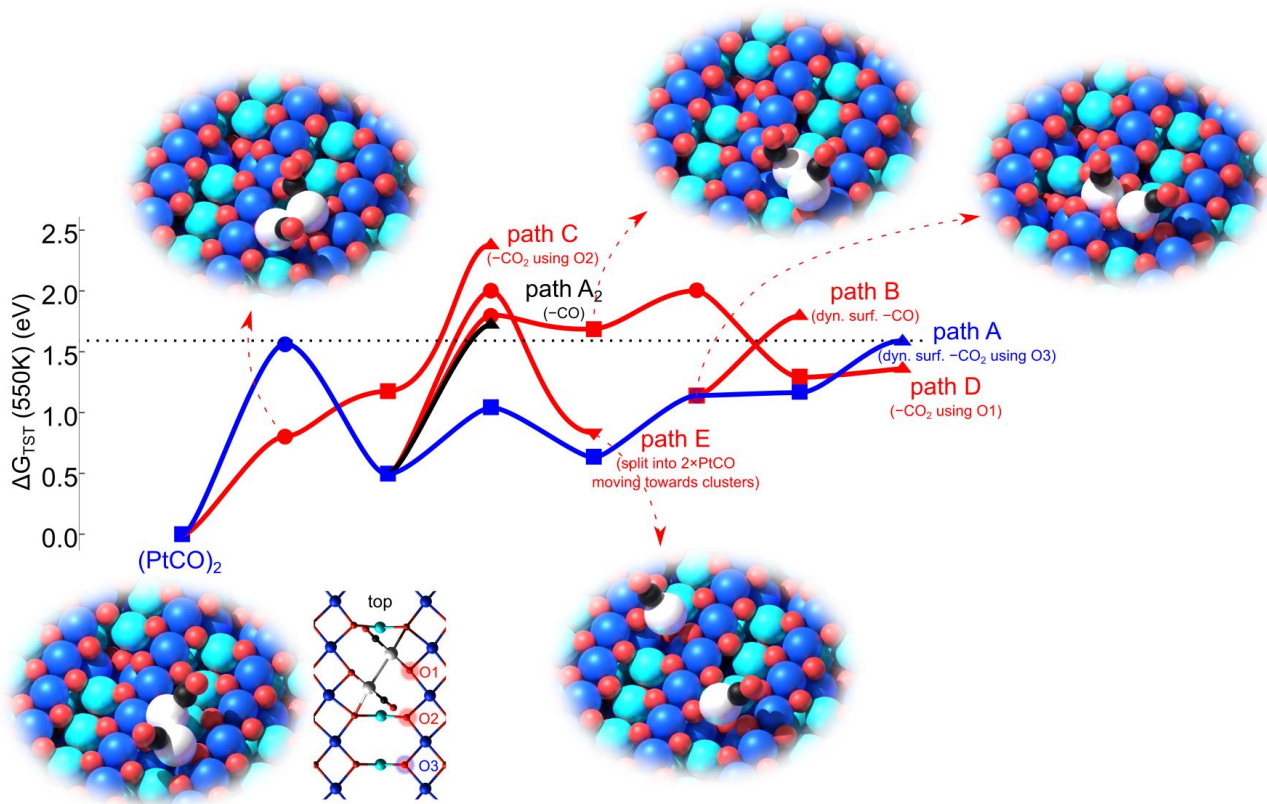

**Figure S3: Alternative (PtCO)₂ pathways (in red, paths B, C, D, E) to the best path leading to CO₂ (blue, path A) or the desorption of CO path (black, path A₂).** Path A and path A2 are described in the main text, and shown in Supplementary Movies path A and path A2. The structures relevant to the red pathways are not shown here, but can be seen in the form of Movies path B, C, D, and E). The dashed line indicates the highest cost of path A. All other paths have either their end-state or a transition state exceeding that cost. Path A₂, direct desorption of CO from the metastable (PtCO)₂ configuration, corresponds to the second most likely path, as discussed in the main text. In this figure the squares represent calculated converged configurations, circles represent cl-NEB climbing image saddle points, and triangle-ups represent desorption processes.

**Path B** proceeds initially in the same way as the favored pathway, but in the final step CO is desorbed rather than CO₂. We speculate that the reason for the relatively low barrier is that the 5-fold coordinated Pt atom would prefer 4-fold coordination to oxygen, as it has in bulk PtO.

**Path C** is based on the stable (PtCO)₂ dimer configuration shown in Fig. 1 in the main text. An adsorbed CO molecule forms a OCO intermediate using the nearest surface oxygen atom, and ultimately desorbs CO₂. There is no intermediate barrier in this process, but the end state is extremely unfavorable. The main reason for this is that the surface oxygen atom in question, O2, has the highest oxygen vacancy formation energy ($V_o$) of all three inequivalent O atoms on the Fe₃O₄(001) surface.

**Path D** begins like the favored pathway with the formation of the metastable (PtCO)₂ dimer. However, in this path, the Fe₃O₄(001) surface remains static. The system distorts the dimer in order to free one of the surface oxygen atoms to which the Pt is bound. The low oxygen vacancy formation energy of these O1 atoms makes the final state lower in energy than that considered in path C, but the process has a significant barrier, and will not occur.

**Path E** corresponds to splitting of the $(PtCO)_2$ dimer. The cause of the high barrier is the strong stability of dimers with respect to PtCO species. Note that this is essentially the reverse of the $(PtCO)_2$ formation pathway shown in the right panel of Fig S2.

### 5- $Pt_2(CO)_1$ (post-CO$_2$-desorption) path

In the main text, the minimum-energy path in Fig. 3 is shown only until the desorption of $CO_2$ from $(PtCO)_2$ species. Movie S4 and Figure S4 show the continuation of this path until the final state is reached. Note that we have reset the energy scale to 0 because the $CO_2$ desorption event is irreversible. This means that the thermal energy at 550 K (circa 1.5 eV) is available for the subsequent steps. As discussed in the main text, here the key result is that the $Pt_2CO$ is species is unstable, and splits into an immobile $Pt_1$ adatom and a PtCO highly mobile at 550 K. The latter species quickly agglomerates with a Pt cluster nucleus and CO desorbs instantly. This explains why we see simultaneous CO and $CO_2$ desorption in TPD, and why the post-TPD state observed in STM is a mixture of adatoms and larger Pt clusters.

Also in this case alternative paths have been considered which turned to be unfavorable and unrealistic and therefore are not reported.

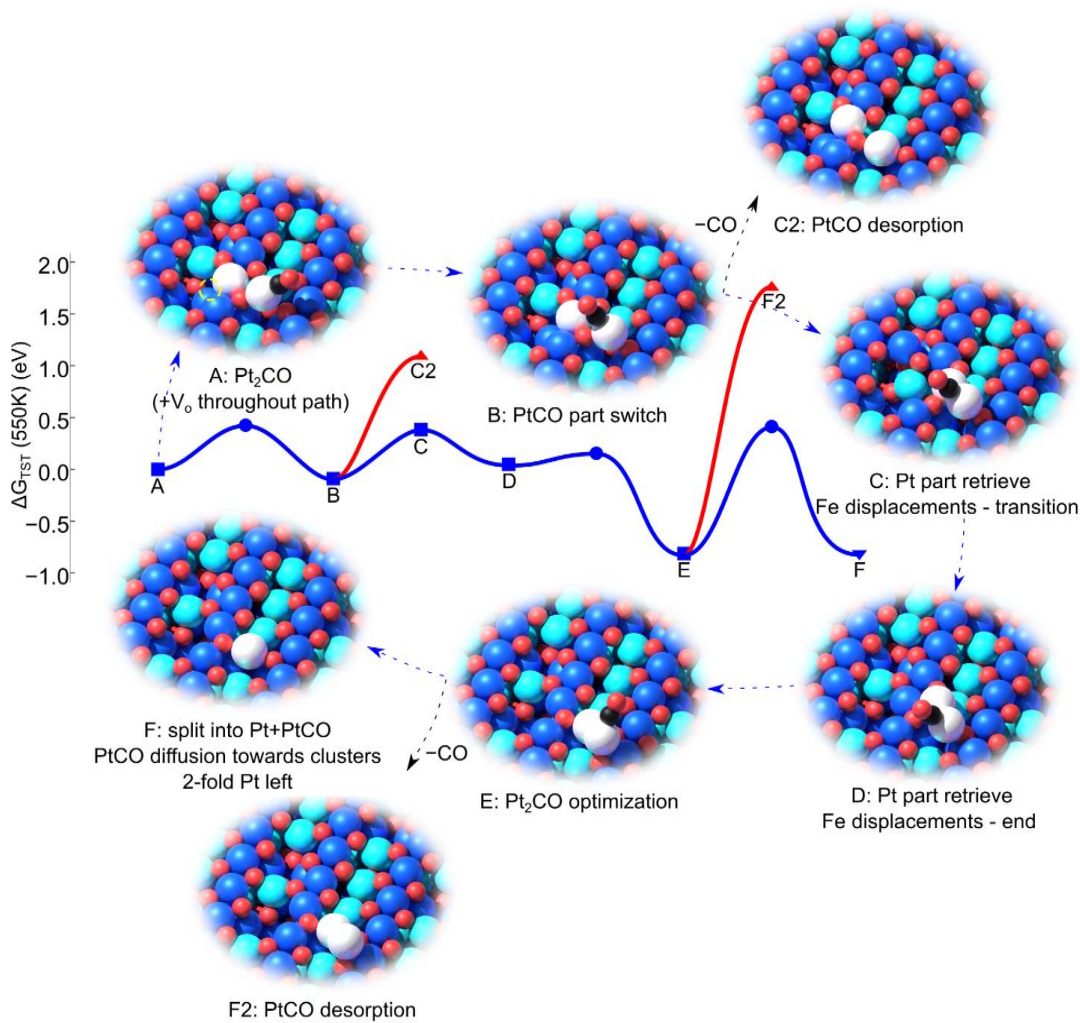

**Figure S4: Post-CO$_2$-desorption minimum-energy path.** The energy scale starts at 0 eV after the desorption of $CO_2$ (an irreversible event), meaning that 1.5 eV is again available for the subsequent processes. The rate-limiting step is the diffusion of the PtCO species on $Fe_3O_4(001)$, which is possible at room temperature, as already shown earlier (see Fig. S2). Note that the different points are shown

together with the animation in Supplementary Movie Path F. The squares represent calculated converged configurations, circles represent cl-NEB climbing image saddle points, and the triangle-down in the last step represents agglomeration.

**Path F**: Immediately after the desorption of $CO_2$, a $Pt_2CO$ entity remains. The previously displaced Fe atoms move to recover the ideal positions within the SCV reconstruction, which now has an oxygen vacancy. The $Pt_2CO$ splits into a mobile PtCO and a 2-fold coordinated $Pt_1$ adatom. Diffusion of the PtCO is easy at this temperature and Pt meets a Pt cluster, with immediate liberation of the adsorbed CO molecule. This last part is not calculated, but it is consistent with both the simultaneous desorption of CO and $CO_2$, as well as the final state, which is a mixture of $Pt_1$ adatoms and Pt clusters.

**Path F2**: The energy required to desorb the CO directly from the $Pt_2CO$ species is 3.37 eV. This process is thus not possible at 550 K.

**Path C2**: The energy required to desorb the CO from state B, where one Pt atom occupies a surface substitutional site and the other a 2-fold adatom site is 1.1 eV. This is significantly less than the barrier for surface diffusion ($\approx$0.5 eV), and thus does not happen before the system sinters to form clusters.

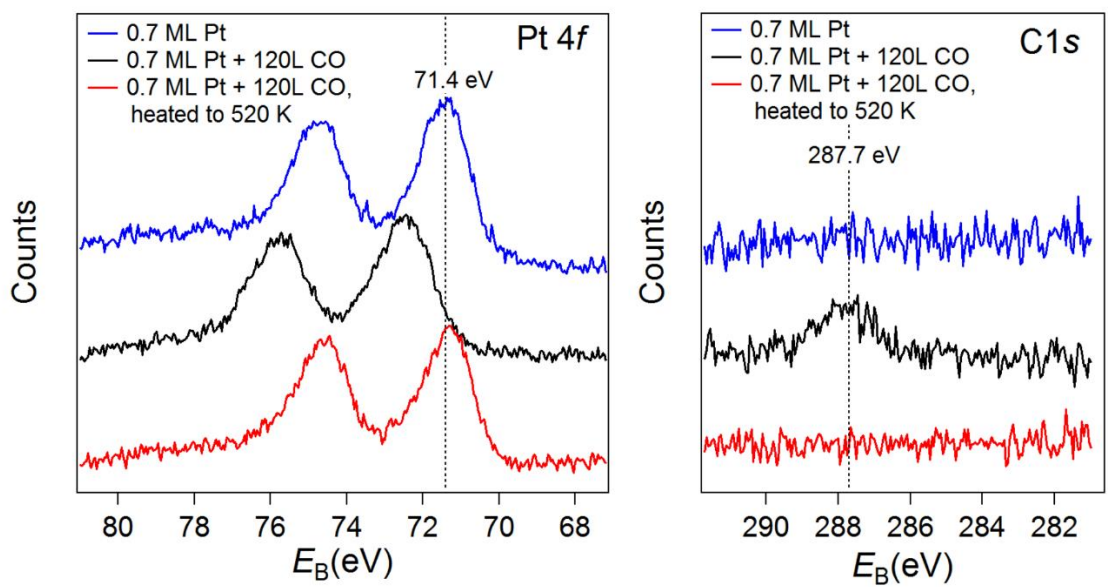

**Figure S5. XPS spectra for the $(PtCO)_2$/$Fe_3O_4(001)$ system acquired before and after CO adsorption, and following CO desorption.** XPS spectra of the Pt 4f and C 1s regions acquired after the deposition of 0.7 ML Pt (blue), after CO exposure at room temperature (20 min, $10^{-7}$ mbar, black), and after annealing to 520 K (red). After CO exposure, CO is visible in C1s, and an upshift of the Pt4f binding energy is observed. Annealing to 520 K causes the CO to desorb, and the Pt peaks shift back to an energy slightly lower than its initial position, closer to that of bulk Pt, ~71.0 eV. Figure reprinted from ref. (39).

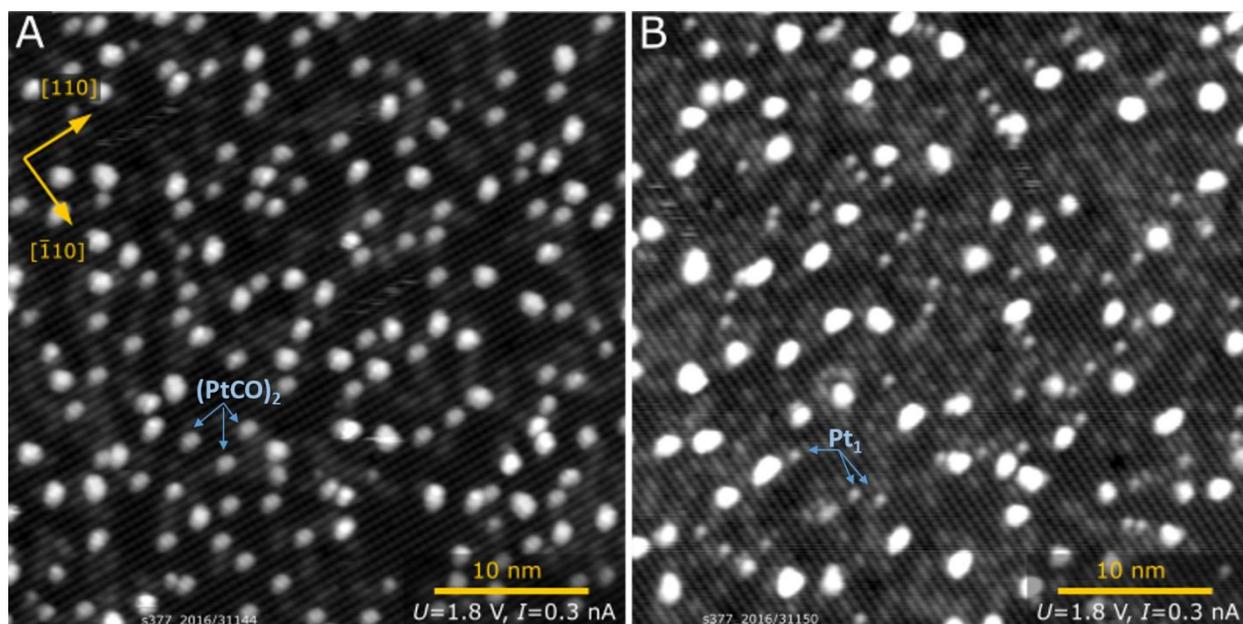

**Figure S6**: **Room-temperature STM images of the Pt/Fe₃O₄(001) system following CO exposure and heating to temperatures (A) 553 K and (B) just above CO/CO₂ desorption.** The image in panel (A) was taken on a surface with 0.2 ML Pt, which was exposed to > 3 L CO and heated to 553 K. Here the $(PtCO)_2$ dimers are clearly present, which indicates the CO and $CO_2$ desorption did not take place yet in the STM setup. The image in panel (B) was taken on the same surface after heating to 583 K. Here the $(PtCO)_2$ dimers are not present anymore, and single $Pt_1$ adatoms and larger $Pt_x$ clusters are observed instead. This image thus clearly shows the surface after CO/CO₂ desorption. Overall, this dataset shows that $(PtCO)_2$ dimers are present on the surface directly before the CO/CO₂ desorption, and thus they are the species most likely responsible for the observed reactivity.

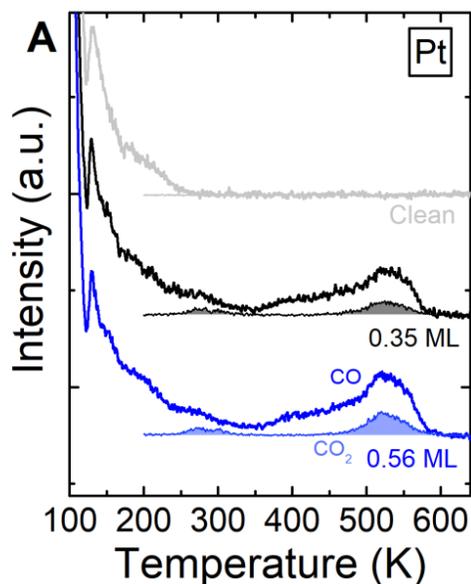

**Figure S7: CO TPD acquired from different coverages of Pt on Fe₃O₄(001).** Thick curves show CO desorption traces, and thin curves with the filled area underneath show CO₂ signals. In each case the surface was exposed to ~3 CO/u.c., which results in the saturation of the surface by CO. Note that CO was dosed at 60 K here, rather than 300 K as in Fig. 2. This does not affect the behavior observed above 300 K.

<u>**Supplementary Movie Captions**</u>

**path A.mp4**: **Short animation showing the minimum energy pathway to form $CO_2$ in the (PtCO)$_2$ system**. The pathway corresponds to the blue curve in Figure 3. In the model, the $Fe_{oct}$ and $Fe_{tet}$ of the $Fe_3O_4(001)$ support are dark blue and cyan, respectively. O atoms are red, Pt are white, and the C and O in CO are black and red, respectively.

**path A2.mp4: Short animation showing an alternative pathway leading to CO desorption from the (PtCO)$_2$ dimer**. The pathway corresponds to the black curve labelled F2 in Figure 3. In the model, the $Fe_{oct}$ and $Fe_{tet}$ of the $Fe_3O_4(001)$ support are dark blue and cyan, respectively. O atoms are red, Pt are white, and the C and O in CO are black and red, respectively.

**path B.mp4: Short animation showing an alternative pathway leading to CO desorption from the (PtCO)$_2$ dimer**. Path B proceeds initially in the same way as the favored pathway, but in the final step CO is desorbed rather than $CO_2$. The pathway corresponds to the red curve labelled Path B in Figure S3. In the model, the $Fe_{oct}$ and $Fe_{tet}$ of the $Fe_3O_4(001)$ support are dark blue and cyan, respectively. O atoms are red, Pt are white, and the C and O in CO are black and red, respectively.

**path C.mp4: Short animation showing alternative pathway Path C, which leads to $CO_2$ desorption from the (PtCO)$_2$ dimer**. In path C, an adsorbed CO molecule forms a OCO intermediate using the nearest surface oxygen atom, and ultimately desorbs $CO_2$. The pathway corresponds to the red curve labelled Path C in Figure S3. In the model, the $Fe_{oct}$ and $Fe_{tet}$ of the $Fe_3O_4(001)$ support are dark blue and cyan, respectively. O atoms are red, Pt are white, and the C and O in CO are black and red, respectively.

**path D.mp4: Short animation showing alternative pathway Path D, which leads to $CO_2$ desorption from the metastable (PtCO)$_2$ dimer**. Path D begins like the favored pathway with the formation of the metastable (PtCO)$_2$ dimer. However, in this path, the $Fe_3O_4(001)$ surface remains static. The system distorts the dimer in order to free one of the surface oxygen atoms to which the Pt is bound. In the model, the $Fe_{oct}$ and $Fe_{tet}$ of the $Fe_3O_4(001)$ support are dark blue and cyan, respectively. O atoms are red, Pt are white, and the C and O in CO are black and red, respectively.

**path E.mp4: Short animation showing an alternative pathway in which the (PtCO)$_2$ dimer splits apart into 2 PtCO species.** The cause of the high barrier is the strong stability of dimers with respect to PtCO species. In the model, the $Fe_{oct}$ and $Fe_{tet}$ of the $Fe_3O_4(001)$ support are dark blue and cyan, respectively. O atoms are red, Pt are white, and the C and O in CO are black and red, respectively.

**path F.mp4: Path relative to Fig. S4 (post-$CO_2$), shown as a short animation.** Immediately after the desorption of $CO_2$, a $Pt_2CO$ entity remains. The previously displaced Fe atoms move to recover the ideal positions within the SCV reconstruction, which now has an oxygen vacancy. The $Pt_2CO$ splits into a mobile PtCO and a 2-fold coordinated $Pt_1$ adatom. Diffusion of the PtCO is easy at this temperature and Pt meets a Pt cluster, with immediate liberation of the adsorbed CO molecule. In the model, the $Fe_{oct}$ and $Fe_{tet}$ of the $Fe_3O_4(001)$ support are dark blue and cyan, respectively. O atoms are red, Pt are white, and the C and O in CO are black and red, respectively.

**Diffusion_PtCO_path-across.mp4: Short animation showing the diffusion of a PtCO across the Fe rows of the $Fe_3O_4(001)$ surface.** In the model, the $Fe_{oct}$ and $Fe_{tet}$ of the $Fe_3O_4(001)$ support are dark blue and cyan, respectively. O atoms are red, Pt are white, and the C and O in CO are black and red, respectively.

**Diffusion_PtCO_path-along.mp4: Short animation showing the diffusion of a PtCO along the Fe rows of the Fe$_3$O$_4$(001) surface.** In the model, the Fe$_{oct}$ and Fe$_{tet}$ of the Fe$_3$O$_4$(001) support are dark blue and cyan, respectively. O atoms are red, Pt are white, and the C and O in CO are black and red, respectively.